\documentstyle[epsfig,here,12pt]{article}
\newcommand{\smallfrac}[2] {\mbox{$\frac{#1}{#2}$}}
\newcommand {\eqref} [1] {(\ref {#1})}
\newcommand {\beq} {\begin{equation}} 
\newcommand {\eeq} {\end{equation}}
 \newcommand {\ber}{\begin{eqnarray*}}
 \newcommand {\eer} {\end{eqnarray*}}
\newcommand {\bea}{\begin{eqnarray}}
 \newcommand {\eea} {\end{eqnarray}} 
\newcommand{\ads}{$AdS_5$ }
\newcommand{\adss}{$AdS_5\times S^5$ }
\def\fin{{\lambda_{\infty}}}
\def\phin{{\Phi_{\infty}}}
\def\tin{{T_{\infty}}}
\begin{document}\begin{titlepage}
\rightline{TAUP-2565-99}
\rightline{\today}
\vskip 1cm
\centerline{{\Large \bf  Confinement in 4D Yang-Mills Theories \newline
    }}
\centerline{{\Large \bf from Non-Critical Type 0 String Theory }}
\vskip 1cm
\centerline{
Adi Armoni\footnote{armoni@post.tau.ac.il},
Ehud Fuchs\footnote{udif@post.tau.ac.il} and
Jacob Sonnenschein\footnote{cobi@post.tau.ac.il}
}
\begin{center}
\em School of Physics and Astronomy
\\Beverly and Raymond Sackler Faculty of Exact Sciences
\\Tel Aviv University, Ramat Aviv, 69978, Israel
\end{center}

\begin{abstract}
We study five dimensional 
non critical type 0 string theory  and its correspondence to non supersymmetric
Yang Mills theory in four dimensions. 
We solve the equations of motion of the low energy effective action
and identify a class of solutions that translates into a confining behavior in the 
IR region of the dual gauge theories. In particular we identify a
setup which is dual to pure $SU(N)$ Yang-Mills theory.
 Possible flows of the solutions to the UV region  are discussed.
 The validity of the solutions and potential
sub-leading string corrections are also discussed.

\end{abstract}
\end{titlepage}

\section{Introduction}

It is well known that the original motivation  behind string theory  
was the search for the theory of the strong interactions. 
In spite of the  fact that string theory ``drifted" to a very different
domain,  
the original quest of a stringy
 description  of  Yang Mills  theory is  still an important challenge. 

A step forward  in this direction was made by 
Polyakov  who  argued \cite{Pol1,polyakov} that this  string theory
is  associated
with a Liouville theory with a  curved fifth dimension.  
Recently, the seminal work of Maldacena\cite{Mal} followed by \cite{GKP}
and \cite{Witten}
invoked a dramatic breakthrough in  the  interplay  
between  string theory and  supersymmetric conformal gauge theory
in the large $N$ limit.
A natural question that has been raised following this development is 
 whether similar insight from gravity and string theory
can be also achieved about  non-conformal and non-supersymmetric
four dimensional gauge theories.  In particular whether one can merge
the ideas of Polyakov with those of the string/gauge duality. 

Indeed  this question attracted  
recently a lot of attention mainly along
the
direction of a critical  type 0 string theory 
\cite{KT1,Minahan1,KT2,gar,KT3,Zarembo,Minahan2,TZ}, but also in a
non-critical
setting  \cite{FM,AG}. A key point in the implementation of a
consistent bosonic string theory
is rendering the tachyon field into a ``good tachyon", namely, shifting
the value of  $m^2$ 
to a positive one.   Klebanov and Tseytlin  
\cite{KT1}  showed that in the background of
D-branes, the coupling  of the tachyon to the 
R-R flux can remove the tachyon instability.
Next it was shown  \cite{Minahan1,KT2} that the UV behavior of
asymptotically free gauge theories can be extracted from the gravity
description of the theories. The Infra-Red behavior of these theories
was first considered in \cite{KT2}, where the IR fixed point of 4d
Yang-Mills with 6 massless adjoint scalars was identified as the
asymptotically $AdS_5\times S^5$ metric. It was then argued
\cite{Minahan2} that the adjoint scalars would acquire mass through
loop corrections and that the generic behavior, in the IR, is that of a pure
Yang-Mills theory. Accordingly, a
confining solution of the gravity equations of motion was found.
Another interesting result   was the
identification of a large $N$ non-supersymmetric CFT\cite{KT3}. The
theory is a $SU(N)\times SU(N)$ gauge theory with 6 adjoint scalars and 4
bifundamental Weyl fermions 
in the $(\bar N,N) \oplus(N,\bar N)$ representation.

In this paper we follow the direction  paved  by Polyakov and   discuss
 non-critical type 0 string theory in five
dimensions. The analysis of the non critical theory faces more
problems than the critical one, however, it has the advantage that 
it does not include degrees of freedoms which are associated with the extra
dimensions.
Hence it is more closely related to the pure YM theory. 
Our main result is that there are  solutions 
to the equations of motion, derived from the effective action,
 that
correspond to confinement in the dual gauge theory.
The notion of confinement here means that a Wilson loop  deduced from 
 a Nambu Goto action in the background of the metric solution admits an
area law behavior.
  The solutions
have non-zero measure in the space of solutions, though other
solutions also exist.
We also derive ``flows" of the solutions that corresponds to 
  the UV behavior of the
gauge theory.  We find
two possible scenarios: (i) Solutions that are connected to an $AdS_5$
solution which corresponds to a UV fixed point. (ii) Solutions in which
the effective gauge coupling is asymptotically free. Unfortunately,
since the analysis is based on numerical integration of differential
equations, we are not able to 
extract the $\beta$ function from the 
 UV behavior of the Wilson loop.
In the absence of flat directions we believe the solutions are associated
with gauge theories that do not include (adjoint) scalars and hence
those gauge theories are in a confining phase. Specifically, 5d type 0
string theory in the background of $N$ D3 ``electric" branes is conjectured
to be
dual to pure 4d Yang-Mills theory\cite{polyakov}. 
We also find certain evidence for a non critical  analog (with no scalars
and one bifundamental fermion) of the $SU(N)\times SU(N)$ gauge theory of \cite{KT3}.
Note that the  notion of ``electric" and ``magnetic" branes 
refers  here to charges with respect to two field strength  five forms.
Obviously,  unlike the critical case  \cite{KT3}, in 5d they are not duals of
each other.   
The coupling of the tachyon
to the R-R forms, which  in this case 
is symmetric under $T\rightarrow -T$,
leads to a confining solution evolving to the UV 
 with a vanishing tachyon.   

Beyond the extraction of solutions admitting ``gauge dynamics",  the purpose of 
our work is to identify  the conditions required to assure the reliability of
 the solutions. We discuss the tadpole cancelation, 
the shift of the tachyonic instability, the consistency of the
non-critical
string, higher order string perturbations, curvature corrections   and the
stability of the stack of $N$ $D3$ branes.

The outline of the paper is as follows.
 We consider the degrees of freedom of the non critical
type 0 theory in five dimensions and  review its low energy effective 
action and  the corresponding equations of motion in section 2.
In section 3  we derive solutions to those equations. An exact \ads
solution,
generic confining solutions and their flows to the UV region.
 In section 4 we discuss  the 
conditions for the validity of the  solutions.
Section 5  is devoted to 
 the field theory interpretation of the gravity solutions.
This is addressed via the symmetries of the boundary theory,  the gauge
coupling,
field content, Wilson loop, 't Hooft loop  and Zig Zag invariance. 
The result of this work are summarized in section 6.
 In the appendix we discuss the asymptotic behavior of the solutions.
 
\section{Non-critical Type 0 String theory}
Let us start with the identification of the low energy degrees of freedom,
namely, the massless (and tachyonic) states of the 5d type 0 string.   
The degrees of freedom in the bulk come from the close
string sectors.

In the 5d non-critical string theory, the degrees of freedom are in the $SO(3)$ representation. From the
$(NS-,NS-)$ sector we get $0\otimes 0=0$ which is the tachyon $T$. From
the $(NS+,NS+)$ sector we get $1\otimes 1=2\oplus 1 \oplus 0$ which
are the graviton $G_{mn}$, the anti-symmetric tensor $B_{mn}$ and
the dilaton $\Phi$.
The R sector vacuum is in the $\frac{1}{2}$ representation of
$Spin(3)\sim SU(2)$.
>From the $(R,R)$ sector we get
$\frac{1}{2}\otimes\frac{1}{2}=1+0$, namely a $0$ and a $1$ forms.
The 1- form of the $(R,R)$ sector is associated
with the 2-form field strength of the $D0$ brane. The 0-form
associated with a 1-form field strength of the $D-1$ brane.
By dualizing the $(p+2)$ field strength of the $Dp$ brane we can get the
$5-(p+2)=(q+2)$ field strength of the $Dq=D(1-p)$ brane. This gives us
the $D1$ and the $D2$ branes. We cannot get the $D3$
branes directly from the R-R sector. Similar thing happens in
10d string theory where one cannot get the $D8$ branes from
the R-R sector. We assume that the $D3$ branes exist because
we can get them from $D2$ branes using T-duality.

The diagonal GSO projection in the critical string includes the states
$(R+,R-)\oplus(R-,R+)$ in type 0A and $(R+,R+)\oplus(R-,R-)$ in type
0B. In the 5d non-critical string there is no chirality and 
hence one cannot assign $+,-$ to the Ramond ground state.
The diagonal GSO projection, does not determine the multiplicity of
the $(R,R)$ sector. However, from the requirement of modular
invariance of the torus partition function, we believe that the (R,R)
sector has to be doubled. 
Assuming there is doubling of the R-R sector we would have two sets of
D-branes, electric and magnetic.

The low energy type 0  effective  action  
built from  the  (NS, NS) and (R,R)   sectors
was derived in \cite{KT1} for any dimension, both critical and
non-critical. We briefly summarize its features using their notations.
The supergravity action from the $(NS+,NS+)$ sector
is the same as the type IIB action
\begin{eqnarray}
S=-2\int d^Dx{\sqrt G}e^{-2\Phi}\left[c_0+R+4(\partial_n\Phi)^2
  -\frac{1}{12}H_{mnk}^2\right] \ ,
\end{eqnarray}
where $c_0=\frac{10-D}{\alpha'}$ is the central charge term that
appears for non critical string theories.
The action of the tachyon that comes from the $(NS-,NS-)$ sector is
\begin{eqnarray}
S=\int d^Dx{\sqrt G}e^{-2\Phi}\left(
  \smallfrac{1}{2}G^{mn}\partial_mT\partial_nT +\smallfrac{1}{2}m^2T^2
  \right) \ ,
\end{eqnarray}
where $m^2=\frac{2-D}{4\alpha'}$ is the mass of the tachyon.
The leading R-R terms in the action are
\begin{eqnarray}
S=\int d^Dx{\sqrt G} h_{(n+1)}^{-1}(T)|F_{n+1}|^2 \ ,
\end{eqnarray}
where $F_{n+1}$ is the field strength of the R-R field and $h_{(n+1)}(T)$
describes the coupling of the tachyon to the R-R field.

For the theory on the background of D-Branes we
use the following ansatz for the metric and the R-R electric field
\begin{eqnarray}
ds^2=d\tau^2+e^{2\lambda(\tau)}dx_\mu^2+e^{2\nu(\tau)}d\Omega_k^2 \ ,\\
C_n=A(\tau)\ ,\ \ \ \ F_{n+1}=A'(\tau) \ ,
\end{eqnarray}
where $x_\mu$ are the coordinates of an $n$ dimensional flat Minkowski
space and $\Omega_k$ is a $k$ dimensional sphere. The string theory is
in $D=n+k+1$ dimensions. $C_n$ is a $n$-form with all the indices in
the Minkowski space, and $F_{n+1}$ is a $(n+1)$-form where the extra
index is in the $\tau$ direction. We also assume for the tachyon that
$T=T(\tau)$.

Defining the function $\varphi\equiv 2\Phi-n\lambda-k\nu$ and plugging the
metric into the string actions we obtain
\begin{eqnarray}
&S=-\int
    d\tau\left(e^{-\varphi}\left[ c_0+k(k-1)e^{-2\nu}-n\lambda'^2-k\nu'^2+\varphi'^2
    - \smallfrac{1}{4}T'^2 - \smallfrac{1}{4}m^2T^2 \right] \right. &\nonumber\\
&   \left. + \smallfrac{1}{4}e^{-n\lambda+k\nu}h^{-1}(T)A'^2 \right) \
\label{action} ,&
\end{eqnarray}

Solving the equation of motion for $A(\tau)$ we obtain
\begin{eqnarray}
A'=2Qe^{n\lambda-k\nu}h(T) \ ,
\end{eqnarray}
where $Q$ is the charge of the R-R field.

In order to have a Toda-like mechanical system we define a new 'time'
parameter $\rho$
\begin{eqnarray}
d\rho=e^\varphi d\tau \ .
\end{eqnarray}

To get a four dimensional gauge theory with no scalars, we
choose $n=4$ and $k=0$. Now the action is independent of $\nu$. We
also set $\alpha ' =1$.
Instead of working with $\varphi$ and $\lambda$, we work with the real
dilaton $\Phi$ and with the function
$\xi=\frac{1}{2}(\varphi+\lambda)$ which
 is the combination of
$\varphi$ and $\lambda$ that diagonalize the action
\begin{eqnarray}
\varphi&=&-\smallfrac{2}{n-1}\Phi+\smallfrac{2n}{n-1}\xi
    =-\smallfrac{2}{3}\Phi+\smallfrac{8}{3}\xi \ ,\\
\lambda&=&\smallfrac{2}{n-1}\Phi-\smallfrac{2}{n-1}\xi
       =\smallfrac{2}{3}\Phi-\smallfrac{2}{3}\xi \ .
\end{eqnarray}

The metric in the new variables is
\begin{eqnarray}
ds^2=e^{\frac{4}{3}\Phi-\frac{16}{3}\xi}d\rho^2
    +e^{\frac{4}{3}\Phi-\frac{4}{3}\xi}dx_\mu^2 \ ,\label{metric}
\end{eqnarray}
and we get the action
\begin{eqnarray}
S&=&\int d\rho \left[ \smallfrac{4}{3}\dot{\Phi}^2
                  - \smallfrac{16}{3}\dot{\xi}^2
                  + \smallfrac{1}{4}\dot{T}^2
                  - V(\Phi,\xi,T) \right] \ ,\\
V(\Phi,\xi,T)&=&(5+\smallfrac{3}{16}T^2)e^{\frac{4}{3}\Phi-\frac{16}{3}\xi}
               -Q^2 h(T)e^{\frac{10}{3}\Phi-\frac{16}{3}\xi}\ .
\end{eqnarray}
The equations of motion from the action are
\begin{eqnarray}
&&\ddot{\Phi}
+ \smallfrac{1}{2}(5+\smallfrac{3}{16}T^2)e^{\frac{4}{3}\Phi-\frac{16}{3}\xi}
- \smallfrac{5}{4}Q^2 h(T)e^{\frac{10}{3}\Phi-\frac{16}{3}\xi}=0 \
\label{phi_eq} ,\\
&&\ddot{\xi}
+ \smallfrac{1}{2}(5+\smallfrac{3}{16}T^2)e^{\frac{4}{3}\Phi-\frac{16}{3}\xi}
- \smallfrac{1}{2}Q^2 h(T)e^{\frac{10}{3}\Phi-\frac{16}{3}\xi}=0 \
\label{xi_eq} ,\\
&&\ddot{T}
+ \smallfrac{3}{4}Te^{\frac{4}{3}\Phi-\frac{16}{3}\xi}
- 2Q^2h'(T)e^{\frac{10}{3}\Phi-\frac{16}{3}\xi}=0 \ .\label{T_eq}
\end{eqnarray}
To derive the full set of equations of motion of the string action
one has to add the zero-energy, ``Gauss-law'' like, constraint
\beq 
\smallfrac{4}{3}\dot{\Phi}^2- \smallfrac{16}{3}\dot{\xi}^2+
\smallfrac{1}{4}\dot{T}^2  + V(\Phi,\xi,T)=0 \ .
\label{Hamiltonian}
\eeq
The function $h(T)$ that describes the coupling of the tachyon to the R-R
field for $N$ parallel D3 electric branes is \cite{KT1}
\begin{eqnarray}
h(T)&=&f^{-1}(T) \\
f(T)&=&1+T+\smallfrac{1}{2}T^2+O(T^3) \approx e^T \ ,
\end{eqnarray}

For the set of electric-magnetic D3 branes, when the electric and
magnetic charges are the same we get \cite{KT3}
\begin{eqnarray}
h(T)=f(T)+f^{-1}(T) \ .\label{h_em}
\end{eqnarray}

Note that the IR behavior of the dual gauge theory is dictated by the
behavior of $\frac{4}{3}(\Phi-\xi)=2\lambda$ which appears in the four
dimensional space-time part of the metric \eqref{metric}.
>From the Wilson loop calculations we know that if $\lambda$ has a
unique minimum at a certain
value of $\rho$ then the theory will confine. Therefore it is useful to
write the metric, the Hamiltonian and the equations of motions in
terms of $\lambda$ and $\Phi$.
The metric is
\beq
 ds^2=e^{-4\Phi+8\lambda}d\rho^2
    +e^{2\lambda}dx_\mu^2 \ . \label{metric2} \\
\eeq
The Hamiltonian is
\bea
H &=& -4\dot \Phi ^2 +16 \dot \Phi \dot \lambda -12 \dot\lambda^2 
      +\smallfrac{1}{4}\dot T^2 \nonumber \\
&&    +(5+\smallfrac{3}{16}T^2) e^{-4\Phi+8\lambda}
      -Q^2 h(T) e^{-2\Phi+8\lambda } =0 \  ,\label{h_lambda}
\eea
and the equations of motion are
\bea
&& \ddot{\Phi}
+ \smallfrac{1}{2}(5+\smallfrac{3}{16}T^2)e^{-4\Phi+8\lambda}
- \smallfrac{5}{4}Q^2 h(T)e^{-2\Phi+8\lambda}=0 \ \label{Phi} ,\\
&&\ddot{\lambda}
- \smallfrac{1}{2}Q^2 h(T)e^{-2\Phi+8\lambda}=0 \label{f} \ ,\\
&&\ddot{T}
+ \smallfrac{3}{4}Te^{-4\Phi+8\lambda}
- 2Q^2h'(T)e^{-2\Phi+8\lambda}=0 \ .\label{T}
\eea

The charge of the R-R field is proportional to the number of branes
$Q\sim N$. It is convenient to absorb this factor by the redefinition
$\Phi = \tilde \Phi - \ln N$, $\lambda =\tilde\lambda - {1\over 2} \ln N$.
This redefinition fixes the $N$ dependence in the metric \eqref{metric2}. The 4d
space-time part is multiplied by ${1\over N}$. Also, the curvature in
the string frame does not depend on $N$. The Einstein frame in 5d in
defined by $ds^2=e^{\frac{4}{3}\Phi}ds_E^2$, therefore the curvature in the
Einstein frame behaves as ${1\over N^{4\over 3}}$.

\section{ Solutions of the equations of motion}

In this section we derive configurations of  the metric, dilaton 
and tachyon  that solve  the 
   equations of motion, and as will be discussed in section 5
incorporate interesting gauge theory interpretation. 
 We find an {\em
  exact} $AdS_5$ solution. 
In addition we  write down solutions 
which later will be argued to  describe the IR regime of
the  corresponding gauge 
theories and we  use numerical analysis to flow to the UV regime.

We discuss both the electric branes system as well as the electric-magnetic
system \cite{KT1,KT3}.
The later is somewhat simpler since in this case $T=0$ is a solution of
the tachyon equation of motion \eqref{T}. This is a consequence of the
symmetry $T\rightarrow -T$ in $h(T)$ which results from the electric-magnetic
symmetry of the problem.

\subsection{The $AdS_5$ solution}

An {\em exact} solution to \eqref{phi_eq} \eqref{xi_eq} and \eqref{T_eq} 
can be found for a general function $h(T)$ by
taking $\Phi$ and $T$ to be constants, and solving the algebraic
equations
\begin{eqnarray}
&&\smallfrac{1}{2}(5+\smallfrac{3}{16}T^2)
- \smallfrac{5}{4}Q^2 h(T)e^{2\Phi}=0 \ ,\\
&&\smallfrac{3}{4}T
- 2Q^2h'(T)e^{2\Phi}=0 \ .
\end{eqnarray}
Then the differential equation for $\xi$ is solved exactly by
$\xi \sim{3\over8}\ln\rho$.
Specifically, for the electric magnetic case \eqref{h_em}, the
solution is
\begin{eqnarray}
&& T = 0 \ , \\
&& \Phi = {1\over 2} \ln {1\over Q^2} \ ,\\
&& \xi = {3\over 8} \ln \rho + {1\over 8} \ln {8\over Q^2} \ ,\label{ads_xi}
\end{eqnarray}
and the metric corresponding to this solution is
\beq
ds^2= {1\over 4\rho ^2} d\rho^2 + {1\over 2^{1\over 2} Q} {1\over \rho
 ^{1\over 2}} dx_{\mu} ^2 \ ,
\eeq
which, upon the replacement $\rho = {1\over u^4}$ can be brought to
the familiar $AdS_5$ form
\beq
ds^2= 4 {du^2 \over u^2} + {1\over 2^{1\over 2} Q} u^2 dx_{\mu} ^2\ , \label{ads}
\eeq
The corresponding curvature is 
\beq
 \alpha '{\cal R} = -5
\eeq
 This solution, which has already appeared 
in \cite{polyakov},
 may seem to have a very  different  curvature  than   that of 
the \adss \cite{Mal}
$\alpha'{\cal R} \sim \sqrt{g_sN}$.  In fact it is of the same nature
since $e^\Phi N\sim 1$. 
Also note that weak string coupling occurs for large $N$.

\subsection{Confining solutions}

Confining gauge theories are characterized in the gravity description
by a unique minimum of the 4d space-time of the metric
\eqref{metric}\cite{KSS}. 
We will elaborate on the gauge theory in section 5. 

We would like to show now that the system of equations
\eqref{h_lambda}-\eqref{T} admits solutions with a minimum.

Assuming $h(T)$ is a positive function, we find using eq.\eqref{f} that
\beq
\ddot{\lambda} > 0 \label{cond}
\eeq 
Hence $\dot\lambda$ is monotonically increasing. Let us assume
boundary conditions such that for small
values of $\rho$, $\dot\lambda$ is negative. Therefore there are two
possibilities: (i) $\dot\lambda$ is negative for all $\rho$.
(ii) $\dot\lambda =0$ at some point. 

A solution in which condition ii. is satisfied
implies  confinement. The reason is that if $\dot\lambda(\rho =
\rho_0)=0$, it is a minimum of $e^{2\lambda}$ and it is guaranteed that
it is a {\em unique} minimum. We demonstrate that such solutions
exist. Moreover, we argue that these solutions are generic.

We find the solutions to the equations of motions by expanding in
power series around the minimum of $\lambda$, assuming $\lambda$ has a
minimum

\begin{eqnarray*}
\tilde \Phi(\rho) &=&  \Sigma _{n=0} ^{\infty} \tilde \Phi_n(\rho-\rho_0)^n \ ,\\
\tilde \lambda(\rho) &=& \Sigma _{n=0} ^{\infty} \tilde \lambda_n(\rho-\rho_0)^n \ , \\
T(\rho) &=& \Sigma _{n=0} ^{\infty} T_n (\rho-\rho_0)^n \ .
\end{eqnarray*}

In order to have a minimum of $\tilde\lambda$ at $\rho=\rho_0$ we set
$\tilde\lambda_1=0$. From the zero energy constraint we get
\begin{eqnarray}
\tilde \Phi_1 &=& \pm\frac{1}{\sqrt{8}}\sqrt{e^{-4\tilde
    \Phi_0+8\tilde\lambda_0}
           \left(5-h(T_0)e^{2\tilde \Phi_0}+\frac{3}{16}T_0^2\right)
         + \frac{1}{2}T_1^2} \ ,\label{Phi_1}
\end{eqnarray}
and from the equations of motion we get
\begin{eqnarray}
\tilde \Phi_2 &=& -\frac{e^{-4\tilde \Phi_0+8\tilde\lambda_0}}{4}
           \left(5-\frac{5}{2}h(T_0)e^{2\tilde
               \Phi_0}+\frac{3}{16}T_0^2\right) \ , \label{solution} \\
\tilde \lambda_2 &=& \smallfrac{1}{4}e^{-2\tilde
  \Phi_0+8\tilde\lambda_0}h(T_0) \ ,\nonumber \\
T_2 &=& -e^{-4\tilde \Phi_0+8\tilde\lambda_0}
        \left(e^{2\tilde \Phi_0}h'(T_0)+\frac{3}{8}T_0\right)\ . \nonumber
\end{eqnarray}
Evidently $\tilde\lambda_2 >0$ and hence the solution corresponds to a
minimum.

 There are five free parameters in the solution:
$\tilde \Phi_0, \tilde \lambda_0, T_0, T_1$ and $\rho_0$.
>From \eqref{Phi_1} we see that some of the parameter space is
excluded by the requirement of a real dilaton.
We can go on and compute higher corrections in $(\rho-\rho_0)$. From
the equation of motion one can see that the $\tilde \Phi_{n+2}, \tilde \lambda_{n+2},
T_{n+2}$ coefficients depend on the $\tilde \Phi_i,\tilde \lambda_i, T_i$ $(i\le n)$
coefficients, meaning that this process can go on indefinitely.

Since the above solutions admit confinement, it is important to
understand that it is not an accidental feature of the solutions.
The three equations of motion have six boundary conditions.
One parameter is decreased by the zero energy constraint. Another one
drops out because the 
Hamiltonian does not depend explicitly on ``time'' and therefore if
$\{\Phi(\rho), \lambda(\rho)\}$ is a solution then $\{\Phi(\rho+\delta
\rho), \lambda(\rho+\delta \rho)\}$ is also a solution. Therefore the physical
space of solutions is
four dimensional. Our solution has four free parameters $\Phi_0,
\lambda_0, T_0$ and $T_1$ and accordingly has a non-zero measure in the space of
solutions.

 The Wilson loop analysis tells us that for large enough 
quark anti-quark distance $L$ the string
``spends most of it's time" near the minimum of $\lambda$
which corresponds to the IR regime.
The question now is how to extrapolate the solutions of above 
to the region  along the $\rho$ direction that corresponds in the gauge
picture to the UV regime.  

\subsection{The flow to the  UV limit}

\begin{figure}[bth]
\centerline{\psfig{figure=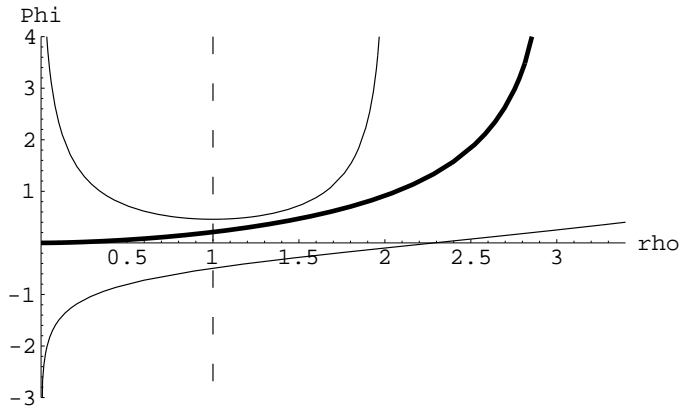,width=9cm,clip=}}
\caption{The dilaton behavior at small distances.
The graph describes 3 kinds of solutions of the dilaton as a function of
$\rho$. The three possibilities are: i. dilaton which flows
to $+\infty$ in the $\rho \rightarrow 0$ limit. ii. $\Phi(\rho=0)=0$
($AdS_5$ solution), which corresponds to a UV fixed point and iii. $\Phi \rightarrow
-\infty$ in the UV which describes asymptotic freedom. The behavior at
$\rho=0$ is dictated by the value of $\Phi$ at the minimum of
$\lambda$ which we draw at $\rho =1$.}
\label{fig:flow}
\end{figure}

To analyze the UV limit we need to continue the solution from the
minimum to the 4d boundary of the 5d space.  
Lets analyze the solution we found for the case of the electric magnetic
branes, $h(T)=e^T+e^{-T}$.
For this function we choose the solutions with $T=0$
to make the analysis easier,
though we found, numerically, similar behavior in the electric case also. 
Now we have only two free parameters $\tilde
\Phi_0,\tilde \lambda_0$
and the solution looks like
\begin{eqnarray}
\tilde \Phi_1 &=& +\frac{1}{\sqrt{8}}\sqrt{e^{-4\tilde
    \Phi_0+8\tilde\lambda_0}
           \left(5-2e^{2\tilde \Phi_0}\right)} \label{Phi_1_2}\\
\tilde \Phi_2 &=& -\smallfrac{1}{4}e^{-4\tilde \Phi_0+8\tilde\lambda_0}
           \left(5-5e^{2\tilde \Phi_0}\right) \\
\tilde \lambda_2 &=& \smallfrac{1}{2}e^{-2\tilde \Phi_0+8\tilde \lambda_0}
\end{eqnarray}
The choice of sign in $\tilde \Phi_1$ is arbitrary. Changing the sign would
give a mirror solution $\rho\rightarrow-\rho$.
>From \eqref{Phi_1_2} we get a constraint on the value of the dilaton
in the minimum of $\lambda$
\begin{eqnarray}
e^{2\tilde \Phi_0} \le \frac{5}{2} \label{confcond}.
\end{eqnarray}
Numerical analysis of the equations of motion shows us that there are
three possible types of behavior for the dilaton, depending on the
choice of parameters $\tilde \Phi_0, \tilde \lambda_0$. If we start with a
large $\tilde \Phi_0$
($e^{2\tilde \Phi_0}$ close to $\frac{5}{2}$), then the dilaton would go to
infinity for $\rho<\rho_0$.
On the other hand, if we start with a small $\tilde \Phi_0$, then the dilaton
would go to minus infinity. Between those two regions in the two
dimensional parameter space there is a border line, at the value
$e^{\tilde \Phi_0} \sim 1.52$, in which the
dilaton $e^{\tilde \Phi}$ flows to $1$ (see figure \ref{fig:flow}).
This set of solutions with the finely
tuned parameters flows to the $AdS_5$ solution \eqref{ads}. 
For the $AdS_5$ solution we know that the boundary is at $\rho=0$ and
$\rho = {1\over u^4}$. Therefore we interpret small values of $\rho$
as large values of energy.

We can compute corrections to the exact $AdS_5$ solution and connect them
to the solution around the minimum. If we plug $\xi$ from \eqref{ads_xi} to the
dilaton equation and assume that the dilaton is small ($e^{\tilde \Phi} \approx
1+\tilde \Phi$) we get
\begin{eqnarray}
4\rho^2\ddot{\tilde \Phi}+5\tilde \Phi=0 \ ,
\end{eqnarray}
which is solved by
\begin{eqnarray}
\tilde \Phi=C \rho^{\frac{1+\sqrt 6}{2}} \label{UVfixed} \ ,
\end{eqnarray}
for any constant $C$. By fixing the value of $C$ we get a $\tilde\Phi$
behavior which is very close to the bold line in figure \ref{fig:flow}
for small $\rho$.
Note that since $\tilde \Phi \rightarrow 0$, $g_{YM}^2N\rightarrow 1$,
we have a UV fixed point at finite value of the 't Hooft coupling.

In order to get asymptotic freedom we need to start with lower values of
$e^{2\tilde \Phi_0}$ in the minimum. Numerical analysis show that those solutions flow toward
$e^{2\tilde \Phi}\rightarrow 0$ in the UV limit.
At the moment we do not know the analytical
behavior of $\Phi$ and $\lambda$.
and therefore we can not extract the $\beta$ function.

\section{Validity of the solutions}

The solutions of the effective Type 0 equations of motion
can be trusted only if certain conditions are obeyed.
Basically one has to check that the  string theory, 
that leads to the low energy effective action whose 
classical configurations we discuss, is consistent.
In addition, higher order  string perturbations and 
sub-leading $\alpha'$ corrections 
should be under control.
On top of these conditions one has to check whether the correspondence
to a ``dual gauge  theory" on the boundary in the spirit of the AdS/CFT
correspondence is consistent. We discuss here the former question and the later
will be addressed in section 5. 
\begin{itemize}
\item The type 0 non critical 
 string theory is consistent only provided that the 
torus partition function is modular invariant and that 
tadpole  of the massless fields are canceled out. 
The critical type 0 string theory, namely the theory derived using a diagonal
GSO projection, was shown to  be modular invariant. We have not performed an
explicit computation in the present non-critical  theory. However
it seems that the same structure of a modular invariant partition function
holds also for our case\cite{Polchinski}. Modular invariance  was argued
also in \cite{polyakov}.

In the type 0 theory there are potentially tachyon and dilaton tadpoles.
The critical type 0 theory discussed in \cite {KT1}
as well as the non-critical one \cite{polyakov} are free from  dilaton tadpoles.
 This is a statement  about the leading behavior of the 
string theory. In fact one has also to assure that  
dilaton tadpoles  are not 
generated even in the sub-leading string corrections\cite{BerRey}.
At present it is not clear to us whether the type 0 models obey this
stronger condition. 

\item  Another necessary condition for the consistency of the
non-critical string theory
is  that  the tachyon $T$ looses its tachyonic nature.
In an AdS background of radius $R$  the requirement is that 
$m^2_T\geq -4/R^2$. For the CFT that is associated with the  critical
dyonic type 0 theory
\cite{KT3}  the condition takes the form $4+16{f'}^2(0)\geq \sqrt{2Q}e^{\Phi/2}$
where $\Phi$ is the value of the dilaton for that model.
In the present case, since we have a non AdS background,
we take the 
more conservative requirement that $m^2_T\geq 0$. Using  \eqref{T_eq}
this translates
into
\begin{eqnarray}
e^{2\tilde \Phi}\geq \frac{3}{8h''(T)} \ ,\label{mt}
\end{eqnarray}
which is obeyed by the solutions that
flow to the $AdS_5$. Note that one has to require the positivity of $m_T^2$ everywhere along the 
$\rho$ coordinate and not only on the surface of the the five
dimensional space. Therefore, the solutions with asymptotic freedom
(small dilaton) do not obey this condition. A solution with asymptotic
freedom can be stable with a non constant tachyon if $h''(T)$ converges in the UV.

\item Non critical string theory  is believed to be  inconsistent
for $c>1$ and $c<c_{crit}$. Our analysis is focused on $d=c=4$. The
obvious
question is whether one can make sense out of such a setup. 
  Polyakov \cite{polyakov} conjectured that  for non flat $d+1$ dimensions,
namely $d$ flat directions and one Liouville direction, the instability of the 
theory may be cured in the presence of non zero R-R background. We
argued above, following \cite{KT1}, that
the mass of  the tachyon can be shifted so that ${m_T}_{eff}>0$
due to the coupling with the R-R field even in the non-critical
dimension of $d=4$. Since the instability of the theories at the 
``forbidden  zone'' past the $c=1$ barrier, manifest itself in the
tachyonic behavior, the shift of $m^2 _T$ may render the theory  at
$d=4$ into a consistent one.  It is important to assure that even
sub-leading string
correction do not introduce other tachyonic modes by shifting the
masses of massless or massive modes to tachyonic ones \cite{BerRey}. 

\item 
The solutions of the equations of motion are reliable only
provided that the modifications of the effective action
due to  higher order string  loop corrections are negligible. For that,
one has to insure that the string coupling $e^\Phi= {1\over N}
e^{\tilde \Phi}$ is weak.
Indeed, the confining solutions which obey \eqref{confcond} and have large
enough $N$, satisfy this condition.

\item 
The gravitational sector of the low energy effective field theory
\eqref{action}  is a valid  approximation to the 
full gravitational  effective action only provided that higher 
order curvature  terms are negligible. 
As was argued in \cite{KT1} the world sheet supersymmetry, even
in the absence of space-time supersymmetry  restricts the
corrections to the type 0 effective action to be identical to those
of the    type II. In  the gravitational sector  the  
first corrections are proportional to ${\alpha'}^3 R^4$, and only
the Weyl tensor contributes to the $R^4$ term. 
It turns out that the five dimensional metric of the general form
\eqref{metric}, not necessarily of an $AdS_5$ form,  
 has a {\it vanishing Weyl tensor} so that the 
first order correction vanishes. 
Since  the full list of higher order curvature corrections is not 
known to us, to be on the safe side, we would like to impose a
restriction that the Ricci scalar is small in the string frame.

The curvature (in the string frame) associated with the solution \eqref{Phi_1}
\eqref{solution} is  given, near the minimum, by 
\begin{eqnarray}
R=-8e^{2\tilde \Phi_0}+O(\rho-\rho_0) \label{curvature}
\end{eqnarray}
Hence, for small $\Phi_0$ we are guaranteed not to have significant
modifications to the gravitational effective action.
Unfortunately, a small $\Phi_0$ is restricted by the requirement for a
stable tachyon \eqref{mt}. We believe that this contradiction appears
because both conditions are too strict. The condition for the small
Ricci scalar is too strict because it seems that the first higher order
curvature corrections depend only on the vanishing Weyl tensor.
The condition for the stable tachyon is too strong because we did not
take into account the fact that the metric
has a negative curvature that helps stabilize the tachyon.

\item  The  stability  of the stack of $N$ $D3$ branes.  
Since there is no  space-time supersymmetry the D branes are not BPS states
and there is a-priori  no reason that there is no force between them. 
Recently,  stable non-BPS states where studied in various setups \cite{sen}.
Even though we believe that the two phenomena  may be related 
the direct implications of \cite{sen} to the present case are still unclear 
to us. 
Note  that  the bosonic degrees of freedom of the type 0 are identical to those
of the type II apart from the fact that in the former case there is a doublet
of R-R fields. 
The computation of the interaction energy  between two parallel $D3$
branes was presented in \cite{KT1} for the case of a single type of $D3$
branes and for electric-magnetic branes.
It seems that the same features show up also in the non-critical type 0 
string. It is therefore our believe that there might be stable stacks
of $N$ electric (or magnetic)  $D3$ brane as well as electric-magnetic
stacks of branes.
Recall that the stability at the level of subleading string
corrections relates to the issue of possible generation of dilaton
tadpole \cite{BerRey}.

In fact the stability in the non-critical case may be in a better
shape than the stability in the critical case, because in the later case
there are flat directions \cite{Zarembo,TZ}.
The 5d non-critical theory has only the Liouville direction perpendicular
to the branes, so there are no flat direction in the first place.   
In the language of the gauge fields this translates into the fact 
that there are no (adjoint) scalar fields and thus the $SU(N)$
gauge symmetry cannot be  spontaneously broken.
 
\item

Unlike the confining solutions  of \cite{Witten2},\cite{BISY}
which  have a ``cigar-like" shape,
the solutions constructed in section 3,
that admit a minimum value  of $e^{2\lambda}$, are characterized generically 
by having two boundaries.
\footnote{ We thank J. Maldacena  for pointing to us this condition}
 The boundary at $\rho=0$ where the 4d field theory
exists as  well as  an additional one at  $\rho\rightarrow \infty$.
In terms of the consistency of the supergravity solution such a scenario is
admissible, however it may be problematic to the dual gauge theory description.
The point is that to guarantee a unitary gauge theory on the boundary
one has to assure that signals cannot propagate in finite time to the other
boundary. This condition translates into $\int_{0}^\infty d\tau e^{-2\lambda(\tau)}
\rightarrow \infty$. To check this condition the metric solution has to be
reliable  not only in the region between $\rho=0$ and $\rho_0$ but also all the 
way to $\rho\rightarrow\infty$. Since toward the later limit the dilaton blows
up we are not able to assure that  our solution obeys the condition.

\end{itemize}

\section{ The gauge theory interpretation}
Now that solutions of the equations of motion were written down and
their validity was considered, we face the challenge of deducing the 
interpretation of the solutions in terms of the four dimensional  
boundary gauge theory. 
Obviously,  in the present non supersymmetric and non conformal scenario 
it is more difficult to argue in favor of 
the duality between the the supergravity solution and
the gauge theory on the boundary.
Nevertheless, we try to analyze  the gauge interpretation
following the recipe of the \adss and the ${\cal N}=4$ SYM correspondence.

\subsection{ Symmetries} 
\begin{itemize}

\item The bulk space-time effective theory of the type 0 string is not
  invariant  under any supersymmetry transformation, therefore the
  "dual" gauge theory should also be a non-supersymmetric one. 

\item The electric solution  and the electric-magnetic solution
  incorporate $N$ ``electric" and  $N$ electric and $N$ magnetic non
  BPS
$D3$ branes respectively.
It is well known that   the open strings between the coinciding $D3$ branes
(of the same type) induce $SU(N)$ gauge fields. Hence the electric theory and the electric-magnetic 
theories should be invariant under local 
 $SU(N)$ and $SU(N)\times SU(N)$ symmetries respectively. 
\item Global symmetries of the boundary  theory originate from isometries 
of the 5d space-time metric (which are in addition to the 3+1 dimensional
Poincare symmetry). The confining solutions do not admit any 
symmetries. The $AdS_5$ has an $SO(2,4)$ isometry group which maps into the
symmetries of the four dimensional conformal gauge theory.   

\end{itemize} 
\subsection{Gauge coupling} 

In the  absence of a tachyon field in the vacuum, the effective gauge theory
on the $D3$ brane has the form of $e^{-\Phi}Tr[F^2]$ so that the gauge coupling
is expected to be $g_{YM} ^2 \sim e^{\Phi}$. It was put forward that this relation translates
the evolution of the coupling constant as a function of the energy scale 
into the  dependence of $\Phi$ on the fifth dimension. In case that
the tachyon is non vanishing in  the ground state, the  gauge
coupling is dressed $e^{\Phi}\rightarrow F(T)e^{\Phi}$. Since one can introduce
a  transformation of the fifth coordinate, an non ambiguous result can be
derived from a ``physical measurement" like the Wilson line discussed
below. Unfortunately, the two prescriptions are in conflict in the
critical type 0 theory\cite{gar}.
Moreover, it happens in the present case even when the tachyon is zero. 

In the UV regime, where the expected behavior of the potential is
${g_{YM}^2(L)N \over L}$, one can read the relation between the gauge
coupling and the dilaton. In the case of critical strings the relation
was $g_{YM}^2 N = e^{{1\over 2}\Phi}$. In the present case it is
$g_{YM}^2 N = e^{{4\over 3}\Phi}$. We will comment about it when we
will discuss the Wilson loop.

\subsection{ Field content}
\begin{itemize}
\item
The full set of fields on the world volume of D branes is determined usually by
the vector fields generated by the open strings, the scalar fields associated
with transverse free motions of the brane and the amount of supersymmetries.
In the present non critical type 0 theory supersymmetry does not 
enforce additional fields in the adjoint representation. Moreover the Liouville
direction is not a free direction and thus a massless scalar field cannot be
associated with  the motion along this direction.
We therefore conclude that the electric theory does not include any  fields apart from 
the gauge fields, namely, it is a pure Yang Mills theory.
In terms of its symmetries, the confining solution and its flow to the UV  is
compatible with such a gauge scenario. 
On the other hand the exact \ads solution dictates a conformal gauge theory. 
This clearly cannot be associate with the quantum Yang Mills theory.
Classically, the theory is conformal invariant but clearly this is not
a property of the full quantum theory.
At present we do not know how to resolve this puzzle. In \cite{polyakov} it is
argued that the \ads solution may be associated with  a UV fixed
point.
But since it seems to be an exact solution for the whole range of $\rho$
this conjecture is not justified.   
\item 
The situation in the electric-magnetic theory is different. As was shown in \cite{Bergman}
the open strings between the electric and magnetic $D$ branes constitute a matter field
on the world volume in the $(N,\bar N) \oplus (\bar N, N)$ representation.
Unlike the conformal models of \cite{KT3} where the global symmetry enforced
a multiplicity of four such representations, in  the present case
since there is no 
$SO(6)$ global symmetry,  there is only one such matter field representation.
The action which includes those fields should not be invariant under any
additional global symmetry.

\end{itemize}

\subsection{The Wilson loop}
A very important tool to translate the super gravity solution into the 
gauge field language is the computation of the Wilson loop introduced in  
\cite{rey,malda2}. By now the Wilson loop and similar objects were computed
in various setups, \cite{Li,BISY}.
In particular in \cite{KSS} a necessary condition
for a confining behavior  was derived. 
We define the functions 
\ber
       f^2(u) & \equiv & G_{tt}(u) G_{xx}(u) \\
       g^2(u) & \equiv & G_{tt}(u) G_{uu}(u) 
\eer
(where $u$ is the fifth  coordinate)
in terms of which the Nambu Goto action for the fundamental string
takes the form

\beq 
      {S_{F1}} =\int d^2 \sigma \sqrt{h} = T \int d\sigma \sqrt{f^2(u) + g^2(u) \: (\partial_x u)^2}
\eeq
 
Then the condition for confinement is that $f$ admits a minimum and the
corresponding string tension is  $\sigma = f(u_{min}) $. 
The setup behind this Nambu-Goto action is such that the quark anti-quark pair
are situated at $u\rightarrow \infty$ and the string connecting them stretches 
in the domain of $\infty>u\geq u_{min}$. In the coordinate assignment of section
2\eqref{metric} the boundary where the external charges are put is at $\rho=0$.
Note also  that in the 
solutions discussed in section 2, $f= e^{2\lambda}$.

In the \ads solution one finds that the quark anti-quark potential takes the
form $ E=c/L$ where $c$ is a numerical factor.

For the \ads fixed point of \eqref{UVfixed}
we are unable to calculate the exact  Wilson loop for
this metric but we can follow the approximation done in \cite{Minahan1}
and assuming $e^{\frac{4}{3}\tilde \Phi}$ is $\rho$ independent for
small $\rho$. Then we get
\begin{eqnarray}
E \sim \frac{e^{\frac{4}{3}\tilde \Phi}}{L} 
\end{eqnarray}
Matching this result to that  in the neighborhood of a UV fixed point
of  a gauge theory namely $E\sim \frac{g_{YM}^2N}{L}$
one can read the relation between the running gauge coupling
and the dilaton.

This is a manifestation of Polyakov's suggestion that the $AdS_5$
solution describes only the extreme UV behavior of the gauge theory
\cite{polyakov}.

Two remarks are in order, (i) the relation between the dilaton  and $g^2_{YM}$
does not agree with the basic relation $e^\Phi\sim g^2_{YM}$ that follows from 
the open string origin of the gauge fields. (ii) It is amusing that for
a non-critical theory in 6 dimensions one does find 
agreement between the naive assignment and the one that follows from the
Wilson loop. 

In the case of asymptotically free solutions, it is very difficult to
extract the the exact behavior of the gauge coupling. The reason is
that the metric in this region is found by numerical integration. In
principle, this is enough to calculate the Wilson loop. Practically,
it is almost hopeless. For the determination of the behavior in this
regime, we will need analytical solutions, which are not at hand at
the moment.   

Next we address the Wilson loop in the IR regime. Though our solution
\eqref{Phi_1} \eqref{solution} is valid at a limited region in the interior of the interval $\rho =
[0,\infty)$, the IR behavior is determined by this regime of the solution.

In particular, the space-time 4d part of the metric \eqref{metric} admits a
minimum at some point $\rho = \rho _0$. As a result \cite{KSS} a confining potential
exists
\beq
V=\sigma r,
\eeq
with the following string tension
\beq
\sigma \sim \min {1\over N} e ^{2\tilde\lambda}  
   ={1\over N}e^{2\tilde\lambda_0} \ .\label{st}
\eeq

It is important to note that since the minimum of $\lambda$ is unique, it is
guaranteed that higher order corrections wouldn't spoil the confining
nature of the solution.

Note that though it seems that the string tension
behaves as \eqref{st} ${1\over N}$ it is not the case. We
may choose the value of $\tilde \lambda_0$ such that $\sigma \sim \Lambda
_{QCD} ^2$ (in $\alpha '=1$ units).

\subsection { The 't Hooft Loop}

Confinement of quarks should be followed by screening of magnetic
monopoles. We would like to check whether our confining
solutions obey this property. We use the recipe of
refs.\cite{Li,BISY}.

The monopole anti-monopole potential is given by the $D1$ action

\beq 
      {S_{D1}} =\int d^2 \sigma e^{-\Phi} \sqrt{h} = T \int d\sigma
      e^{-\Phi} \sqrt{f^2(u) +  g^2(u) \: (\partial_x u)^2}
\eeq

Thus the question of screening versus confinement in this case depends
on the behavior of $e^{-\Phi} f = e^{-\Phi +2\lambda}$. Though we can't prove that
magnetic confinement is excluded, the numerical behavior of $\Phi$ and
$\lambda$ tells that
for quark confining solutions which flow to UV fixed point or to UV
free behavior, we {\em do have} magnetic screening.

\subsection{ Zig Zag invariance}
Polyakov \cite{polyakov} stated the invariance of the Wilson loop
under the Zig Zag transformation as a
necessary condition for any  string  theory description of a gauge theory.
The condition translates to the following requirement on the metric
\beq
e^{\lambda(\tau_*)}=0\ or\  \infty
\eeq
where $\tau_*$ is the value of $\tau$ at the boundary field theory.
 This condition follows from the fact that the solutions found
for $\Phi$ and $\lambda$ do not extrimize the boundary term  in the effective
action\cite{polyakov} and thus the $D3$ branes are driven to either $\tau=0$ or
$\tau=\infty$. Polyakov further advocated the option of
$e^{\lambda(\tau_*)}= \infty$ for which massive boundary states vanish,
and the momentum of massless ones is not restricted.
The only massless states should be the gauge fields. As argued in 5.3 indeed
there are no additional massless
 scalar fields on the boundary and in the electric case no additional fields at
all.

The \ads solutions, both the exact (section 3.1)
 and the UV fixed point (section 3.3),
obey this condition since for those cases $\lambda\rightarrow \infty$.
The flow  of the confining solution to  a solution with an ``asymptotic
freedom" does not obey this condition since  $e^{\lambda{\tau_*}}$ is a constant.

\section{Summary and Discussion}

Whereas the duality between  4d SYM in the large $N$ limit and the
string theory on an \adss background is by now well established, 
there are only first hints that a similar correspondence may
be applicable also to the pure YM theory. Moreover, it seems  now that the 
more promising avenue to construct a string theory of strong interaction
is via a gauge/gravity  duality.
Our work is aimed at improving the understanding  of this approach. 

Motivated by the  required form of the space time metric  that yields
confining Wilson loop \cite{KSS} we searched for solutions of the equations of motion
for which $\lambda$ has a minimum. 
A class of such solutions was identified.  We showed also that the 
solution around the region of the minimum is smoothly connected to the 
region of   small
values of $\rho$ and found that in the region of small $e^{2\tilde
  \Phi_0}$, $\tilde \Phi \rightarrow -\infty$. Therefore, the 
corresponding gauge theories are UV free.

It is interesting to check whether the flow towards the UV
is logarithmic as a function of the energy scale. While in other
similar cases \cite{Minahan1,KT2,Minahan2} it was shown that this is
indeed the case, we haven't find such analytical solutions in the
present case.  Nevertheless, the numerical study does not exclude such
a possibility. The analysis was performed in the ``electric-magnetic
theory'' only, where $T=0$. It is important to analyze the UV regime
in the electric theory which describes pure Yang-Mills. We postpone
these issues for the future.
Another interesting direction is to extract the glueballs mass spectrum and
to compare it to lattice results. For this one has to know the full
behavior of the metric.

A point that was not emphasized enough in the study of the type 0 string 
is that the large $N$ limit  seems to be different then the one
implemented in the original duality of Maldacena \cite{Mal}.
To guarantee the consistency of the solutions, namely small string coupling and
small curvature large $N$ was useful but not large $g_sN$. In fact we found that
what was needed is $g_sN>3/16$ to guarantee the removal of the tachyonic behavior
and $g_sN<1$ to assure small scalar curvature in the string frame.

On the route to the solutions several assumptions were made which deserves 
further justification. In particular  an explicit
computation of the torus partition function  has to be performed
so that modular invariance could be checked. Additional checks about
possible generation of tadpoles of massless particles due to sub leading 
correction are also needed.

An interesting issue which has not been investigated thoroughly enough is
the relation between  the type 0 $D$ branes that were used in the present 
work (and also in  the studies made in the critical setting)  and the 
the stable non BPS D branes discussed in \cite{sen}. In particular 
It seems to us that the 
non trivial tachyon profiles discovered recently \cite{sen2} 
may play an important role also in the string solutions that corresponds to 
gauge dynamics. 

Recently, several authors discussed super-gravity solutions of type IIB,
in which the matter fields are massive\cite{KS,gubser,gppz}. As expected from
the field theory side, these solutions also admits confinement. It
would be interesting to compare their results to ours.

\section{Acknowledgments}
We thank Micha Berkooz, Nissan Itzhaki and Barak Kol for useful
discussions. We would specially like to thank Shimon Yankielowicz for many
useful and fruitful discussions.
The work of J.S. is supported in part by the
US-Israrel Binational Science Foundation, by GIF - the German-Israeli
Foundation for Science Research, and by the Israel Science Foundation.

\section{Appendix - Asymptotic behavior}
While it is difficult to find the precise behavior of the solutions at
large $\rho$ we can negate some possibilities. Since $\dot\lambda$ is
monotonically increasing, it might approach some limit at
$\rho\rightarrow \infty$. We would like to show that while it is impossible in the
theory which lives on the electric-magnetic sets of branes, there is a
room for such scenario in the electric branes theory (pure
Yang-Mills).

Let us first assume zero tachyon solution with the following large
$\rho$ behavior
\bea
&& \lambda \rightarrow \fin \rho \ , \\
&& \Phi \rightarrow \phin \rho \ . 
\eea    
For the fall-off of the exponentials in \eqref{Phi}\eqref{f} we impose
\bea
&&  -2\phin + 8 \fin <0 \ ,\\
&& -4\phin + 8 \fin <0 \ .
\eea
In addition
\beq
 \fin >0 \ ,
\eeq
and from the Hamiltonian
\beq
-4\phin ^2 + 16 \phin \fin - 12 \fin ^2 =0 \ ,
\eeq 
which cannot be satisfied with the above constraints.

In the case of non-zero tachyon and $h(T)=e^{-T}$ we may assume a
similar asymptotic behavior for the tachyon
\beq
T \rightarrow \tin \rho \ .
\eeq
Now, the requirements from the exponentials are
 \bea
&&  -\tin -2\phin + 8 \fin <0 \ ,\\
&& -4\phin + 8 \fin <0 \ .
\eea
and from the Hamiltonian
\beq
-4\phin ^2 + 16 \phin \fin - 12 \fin ^2 + {1\over 4} \tin ^2=0 \ .
\eeq
These conditions are satisfied in the range ${1\over 4} \phin > \fin >0$.

Therefore when a tachyon is included, the asymptotic behavior of
$\Phi, \lambda$ and $T$ can be linear at large values of $\rho$, while when
$T=0$ (and $h(T)$ is an even function) the derivative of $\Phi$ or $\lambda$
must diverge.

\end{document}